\documentstyle[12pt,epsf,epsfig]{article}

\newlength{\extraspace}
\newlength{\extraspaces}
\newcommand{\be}{\begin{equation}}
\newcommand{\ee}{\end{equation}}
\newcommand{\bea}{\begin{eqnarray}}
\newcommand{\nn}{\nonumber}
\newcommand{\eea}{\end{eqnarray}}
\newcommand{\nd}[1]{/\hspace{-0.6em} #1}
\newcommand{\nk}{\noindent}

\def\lsim{\mathrel{\rlap {\raise.5ex\hbox{$ < $}}
{\lower.5ex\hbox{$\sim$}}}}

\def\gappeq{\mathrel{\rlap {\raise.5ex\hbox{$>$}}
{\lower.5ex\hbox{$\sim$}}}}
\def\lappeq{\mathrel{\rlap{\raise.5ex\hbox{$<$}}
{\lower.5ex\hbox{$\sim$}}}}

\begin{document}

\begin{titlepage}
\begin{flushright}
hep-th/9704169 \\
CERN-TH/97-58 \\
CTP-TAMU-22/97 \\
ACT-07/97 \\
OUTP-97--16P \\
\end{flushright}
\begin{centering}
\vspace{.1in}
{\large {\bf Quantum
Decoherence in a $D$-Foam Background }} \\
\vspace{.2in}
{\bf John Ellis$^{a} $},
{\bf N.E. Mavromatos$^{b,\diamond}$},
{\bf D.V. Nanopoulos}$^{c,d,e}$ \\
\vspace{.03in}
\vspace{.1in}
{\bf Abstract} \\

Within the general framework of Liouville string theory, we
construct a model for quantum $D$-brane fluctuations in the space-time
background through which light closed-string states propagate.
The model is based on monopole and vortex defects on the world sheet,
which have been discussed previously in a treatment of $1+1$-dimensional
black-hole fluctuations in the space-time background, and makes use of
a $T$-duality transformation to relate formulations with Neumann and
Dirichlet boundary conditions. In accordance with previous general
arguments, we derive an open quantum-mechanical description of this
$D$-brane foam which embodies momentum and energy conservation and small
mean energy fluctuations. Quantum decoherence effects appear at a rate
consistent with previous estimates.
\vspace{.05in}
\end{centering}
{\small  

}
\vspace{0.2in}
\nk $^a$ Theory Division, CERN, CH-1211, Geneva, Switzerland,  \\
$^b$ University of Oxford, Dept. of Physics
(Theoretical Physics),
1 Keble Road, Oxford OX1 3NP, United Kingdom,   \\
$^{c}$ Center for
Theoretical Physics, Dept. of Physics,
Texas A \& M University, College Station, TX 77843-4242, USA, \\
$^{d}$ Astroparticle Physics Group, Houston
Advanced Research Center (HARC), The Mitchell Campus,
Woodlands, TX 77381, USA. \\
$^{e}$ Academy of Athens, Chair of Theoretical Physics, 
Division of Natural Sciences, 28 Panepistimiou Avenue, 
Athens 10679, Greece. \\
$^{\diamond}$ P.P.A.R.C. Advanced Fellow.

\vspace{0.01in}
\begin{flushleft}
CERN-TH/97-58 \\
CTP-TAMU-22/97 \\
ACT-07/97 \\
OUTP-97-16P \\
April  1997\\
\end{flushleft}
\end{titlepage}
\newpage


Our objective in this paper is to set up a formalism suitable
for describing the propagation of a low-energy light particle,
interpreted as a closed string state, through a fluctuating 
quantum space-time background, for which we adapt and develop 
emergent $D$-brane technology. The physical question towards which 
this study is directed is whether conventional quantum mechanics
can be maintained in the presence of such a space-time foam.
Hawking has argued~\cite{hawk} that a quantum state propagating 
through such a fluctuating background inevitably decoheres in general, 
as a result of information loss through Planck-scale event horizons. 
His arguments were based heuristically on model calculations in a 
quantum treatment of a conventional field-theoretical 
gravity~\cite{hawk,ehns}.  
We have made similar arguments~\cite{emn} and developed 
an appropriate open quantum-mechanical treatment of the 
evolution of a microscopic system, using as a guide 
a $(1+1)$-dimensional string black-hole model~\cite{witt}. 

We argued that, whilst the Hawking-Bekenstein 
black-hole entropy was given by the 
number of stringy black-hole microstates in $(1+1)$ and presumably in $(3+1)$
dimensions~\cite{EMNcount}, and these were in principle distinguishable,
enabling quantum coherence to be maintained 
at the full string level, {\it in practice} experiments
{\it do not} measure {\it all} these microstates, and hence entanglement 
entropy grows and effective quantum coherence is lost. Formally, we argued
that this physics could be represented by Liouville string~\cite{aben,ddk}, 
with 
microscopic black holes in the background space-time
foam pushing the theory away from criticality, leading to 
non-trivial dynamics for the Liouville field~\cite{emn}, which should be interpreted 
as the time variable~\cite{emn,aben,kogan}. 

A more complete formalism for higher-dimensional black holes
in string theory has now been provided by $D$ branes~\cite{dbranes}, 
whose quantum states give an exact microscopic accounting for the 
black-hole entropy in higher dimensions, 
providing a laboratory for the explicit 
extension of our previous arguments on quantum decoherence to the realistic 
$(3+1)$-dimensional case. As a first step in this programme, we have 
demonstrated~\cite{diffusion} that quantum recoil effects in the scattering 
of a light closed-string state off a $D$ brane generate entanglement
entropy 
in the light-particle system when one sums over the unseen 
quantum excitations of the recoiling $D$ brane. 

The next step, undertaken in this paper, is the treatment of 
quantum $D$-brane 
fluctuations in the space-time background, and the propagation 
of light-particle states through this $D$-brane foam. 
To this end, we first recall relevant aspects of our $(1+1)$-dimensional 
string black-hole analysis~\cite{emn}, 
discussing correlators in Liouville theory, and showing 
that S-matrix elements are not in general well defined, whereas $\nd{S}$-matrix 
elements are. As an analogue for this technical development, we make
an explicit 
connection between our Liouville formalism and the closed-time path (CTP)
formalism~\cite{ctp} used in conventional finite-density 
field theory. Next we show that the appearance and disappearance 
of virtual $D$ branes may be represented by monopole-antimonopole pairs 
on the world-sheet, as was previously shown~\cite{emn} to be the case 
in the $(1+1)$-dimensional string model. Monopole-antimonopole 
pairs are 
connected by Dirac-string
singularities which cut {\it slits} along the world sheet, introducing 
open strings, which may be given Dirichlet boundary conditions. 
This formalism provides an explicit realization 
of the Liouville-string approach, including the  departure from criticality 
and the resulting non-trivial Liouville dynamics, from which we derive 
a time-evolution equation for the effective light-state density matrix that is 
reminiscent of open quantum-mechanical systems and incorporates quantum decoherence. 

To establish our basic framework, we first
consider a generic conformal field theory action $S[g^*]$ perturbed by a 
non-conformal deformation $\int d^2z g^iV_i $, whose couplings 
have world-sheet renormalization-group $\beta$-functions $\beta_i=(h_i-2)(g^i-g^{*i})
+ c^i_{jk}(g^j-g^{*j})(g^k-g^{*k}) + \dots $, 
where the $c^i_{jk}$ are operator product expansion 
(OPE) coefficients defined in the normal way. Coupling this theory to 
two-dimensional quantum gravity restores conformal invariance at the quantum 
level, by introducing the Liouville mode $\phi$, which scales the world-sheet
metric $\gamma_{\alpha\beta}=e^{\varphi}{\hat \gamma}_{\alpha\beta}
\equiv e^{\phi/Q}{\hat \gamma}_{\alpha\beta}$, 
with $Q$ to be defined below,  
and ${\hat \gamma}$ a some suitable 
fiducial metric, and makes the gravitationally-dressed operators $[V_i]_\phi$ 
exactly marginal. The corresponding 
gravitationally-dressed conformal theory is:
\be
S_{L-m} = S[g^*] + \frac{1}{4\pi\alpha '}
\int d^2 z \{\partial _\alpha \phi \partial ^\alpha \phi
- QR^{(2)} + \lambda ^i(\phi ) V_i \}
\label{C5}
\ee
where~\cite{ddk,polkle}:
\be
\lambda ^i(\phi ) =g^i e^{\alpha _i \phi }
+ \frac{\pi}{Q \pm 2\alpha _i } c^i_{jk} g^jg^k
\phi e^{\alpha _i \phi } + \dots~~;~~\alpha _i = -\frac{Q}{2} +
\sqrt{\frac{Q^2}{4} - (h_i - 2)}
\label{C6}
\ee
We identify the Liouville field $\phi$ with a dynamical 
local scale on the world-sheet and focus 
on the operators $V_i$ that are $(1,1)$ but not exactly marginal,
i.e. have $h_i=2$ but $c^i_{jk} \ne 0$. The couplings obey 
$\frac{d}{d \tau}\lambda^i(\phi)=\beta^i$, where 
$\tau=-\frac{1}{\alpha Q}{\rm ln}A:~A\equiv \int d^2z \sqrt{{\hat \gamma}}
e^{\alpha\phi(z,{\bar z})}$ is the world-sheet area, and 
$\alpha = -\frac{Q}{2}+\frac{1}{2}\sqrt{Q^2+8}$ with:
\be
Q=\sqrt{\frac{|25-C[g,\phi]|}{3g_s^\chi} + \frac{1}{2}\beta^iG_{ij}\beta^j}
\label{31/2}
\ee
Here $C[g,\phi]$ is the Zamolodchikov $C$  function~\cite{zam},
which 
on account 
of the $C$ theorem~\cite{zam} is given by: 
\be  
  C[g, \phi]=c^*-g_s^\chi \int_{\phi_*}^{\phi}d\phi'\beta^iG_{ij}\beta^j
\label{ctheorem}
\ee 
where $*$
denotes a fixed point of the world-sheet flow,
$g_s=e^{-<\Phi>}$ is the string coupling~\footnote{The 
explicit powers of 
the string coupling constant 
are due to the fact that 
in string theory the 
$C$ function is a target-space-time integral over a measure
$\int d^DX\sqrt{G}e^{<\Phi> \chi}$, where $G$ is the target space 
metric.
Such normalization factors accompany any $\sigma$-model vacuum
expectation value $< \dots >$.}, $\Phi$ is the dilaton field, 
$\chi$ is the Euler characteristic of the world-sheet manifold, 
and the other part of $Q$ is due to the local character of the 
renormalization-group scale~\cite{emn}, with $G_{ij}$ related to divergences 
of $<V_iV_j>$ and hence the Zamolodchikov metric~\cite{zam}.

Correlation functions in such a theory may be written in the form
\be
A_N \equiv <V_{i_1} \dots V_{i_N} >_\mu = \Gamma (-s) \mu ^s
<(\int d^2z \sqrt{{\hat \gamma }}e^{\alpha \phi })^s {\tilde
V}_{i_1} \dots {\tilde V}_{i_N} >_{\mu =0}
\label{C12}
\ee
where the ${\tilde V}_i$ 
have the Liouville zero mode removed, $\mu$ is a 
scale related to the world-sheet cosmological constant, 
and $s$ is the sum of the anomalous dimensions of the 
$V_i~:~s=-\sum _{i=1}^{N} \frac{\alpha _i}{\alpha } - \frac{Q}{\alpha}$.
As it stands, (\ref{C12}) 
is ill defined for $s=n^+ \in Z^+$, because of the 
$\Gamma (-s)$ factor~\cite{emndollar}. 
To regularize this factor, we use the 
integral representation~\cite{kogan2,emn} 
$\Gamma (-s)=\int dA e^{-A} A^{-s-1}$, 
where $A$ is the covariant area of the world sheet, 
and analytically continue to the contour shown in 
Fig. 1. Intepreting the Liouville field $\phi$ 
as time~\cite{emn}: $t \propto {\rm ln}A$, we interpret 
the contour of Fig. 1 as representing evolution 
in both directions of time between fixed points of the 
renormalization group: $ {\rm Infrared} ~ {\rm fixed} ~ 
{\rm point}  \rightarrow {\rm  Ultraviolet} ~ {\rm fixed}
~{\rm point} \rightarrow
 {\rm Infrared} ~ {\rm fixed} ~ {\rm point}$.

\begin{figure}
\hglue3.0cm
\epsfig{figure=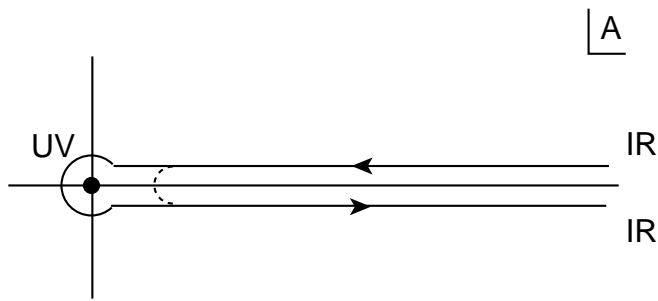}
\caption{ The solid line is the       
the Saalschutz contour in the complex
area ($A$) plane, which is used to 
continue analytically the prefactor 
$\Gamma (-s)$ for $ s \in Z^+$; 
it has been used in
conventional quantum field theory to relate dimensional
regularization to the Bogoliubov-Parasiuk-Hepp-Zimmermann
renormalization method. The dashed line denotes the 
regularized contour, which avoids the ultraviolet 
fixed point $A \rightarrow 0$, which is used in the Closed Time-like Path 
formalism.}
\end{figure}

Within this approach, it is not difficult to see that conventional 
$S$-matrix elements are 
in general ill-defined in Liouville-string theory, and that
scattering 
must be described by a non-factorizable $\nd{S}$-matrix. 
Decomposing the Liouville field in an orthonormal 
mode sum: 
$\phi (z, {\bar z}) = \sum _{n} c_n \phi _n  = c_0 \phi _0
 + \sum _{n \ne 0} \phi _n$, 
where $   \nabla ^2 \phi _n = -\epsilon_n \phi _n~:~n=0, 1,2, \dots$, and
we separate the zero mode 
$\phi _0 \propto A^{-\frac{1}{2}}$
with $\epsilon _0 =0$. The correlation 
function with $\phi_0$ subtracted may be written as 
\bea
{\tilde A}_N \propto &\int & \Pi _{n\ne0}dc_n exp(-\frac{1}{8\pi}
\sum _{n\ne 0} \epsilon _n c_n^2 - \frac{Q}{8\pi}
\sum _{n\ne 0} R_n c_n + \nn \\
~&~&\sum _{n\ne 0}\alpha _i \phi _n (z_i) c_n )(\int d^2\xi
\sqrt{{\hat \gamma }}e^{\alpha\sum _{n\ne 0}\phi _n c_n } )^s
\label{C16}
\eea
with $R_n = \int d^2\xi R^{(2)}(\xi )\phi _n $. 
We can compute
(\ref{C16}) by analytically continuing \cite{goulian}
$s$ to a positive integer $s \rightarrow n^{+} \in {\bf Z}^{+} $.
Denoting
$f(x,y) \equiv  \sum _{n,m~\ne 0} \frac{\phi _n (x) \phi _m (y)}
{\epsilon _n}$
and integrating over the $c_n$, we find 
\bea
~&& {\tilde A}_{n + N} \propto
exp[\frac{1}{2} \sum _{i,j} \alpha _i \alpha _j
f(z_i,z_j) + \nn  \\
~&&\frac{Q^2}{128\pi^2}
\int \int  R(x)R(y)f(x,y) - \sum _{i} \frac{Q}{8\pi}
\alpha _i \int \sqrt{{\hat \gamma}} R(x) f(x,z_i) ]
\label{C19}
\eea
When one makes an infinitesimal Weyl transformation  
$\gamma (x,y) \rightarrow \gamma (x,y) ( 1 - \sigma (x, y))$,
the correlator ${\tilde A}_N$
transforms as follows~\cite{mn,lag}:
\bea
&~&
\delta {\tilde A}_N \propto
[\sum _i h_i \sigma (z_i ) + \frac{Q^2}{16 \pi }
\int d^2x \sqrt{{\hat \gamma }} {\hat R} \sigma (x) +    \nn \\
&~&
\frac{1}{{\hat A}} \{
Qs \int d^2x \sqrt{{\hat \gamma }} \sigma (x)
       +
(s)^2 \int d^2x \sqrt{{\hat \gamma }} \sigma (x) {\hat f}_R (x,x)
+  \nn \\
&~&
Qs \int \int d^2x d^2y
\sqrt{{\hat \gamma }} R (x) \sigma (y) {\hat {\cal
 G}} (x,y) -
  s \sum _i \alpha _i
  \int d^2x
  \sqrt{{\hat \gamma }} \sigma (x) {\hat {\cal
 G}} (x, z_i) -   \nn \\
&~&
 \frac{1}{2} s \sum _i \alpha _i{\hat f}_R (z_i, z_i )
  \int d^2x \sqrt{{\hat \gamma }} \sigma (x)
-    \nn \\
&~&
 \frac{Qs}{16\pi} \int
  \int d^2x d^2y \sqrt{{\hat \gamma (x)}{\hat \gamma }(y)}
  {\hat R}(x) {\hat f}_R (x,x) \sigma (y)\} ] {\tilde A }_N
\label{dollar}
\eea
where the hat notation denotes transformed quantities,
and
$  {\cal G}(z,\omega ) \equiv
f(z,\omega ) -\frac{1}{2} (f_R (z,z) + f_R (\omega, \omega ) )$:
$f_R(z,z)={\rm lim}_{\omega \rightarrow z}\left(f(z,\omega)
+ {\rm ln}d^2(z,\omega)\right) $, where $d(z,\omega)$ 
is the geodesic distance on the world sheet. 
We see explicitly that (\ref{dollar}) contains non-covariant 
terms $\propto A^{-1}$ 
if the sum of the anomalous dimensions
$s \ne 0$. Thus the generic correlation 
function $A_N$ does not have a well-defined 
limit as $A \rightarrow 0$.

In \cite{emn} 
we identified the target time as $t=\phi_0=-{\rm ln}A$, 
where $\phi_0$ is the world-sheet zero mode of the Liouville field.
The normalization follows from
a consequence 
of the canonical form of the kinetic term for the Liouville field $\phi$ 
in the Liouville $\sigma$ model~\cite{aben,emn}. 
The opposite flow of the target time, as compared to that of the 
Liouville mode, is, on the other hand, a consequence
of the `bounce' picture~\cite{kogan2,emn} for Liouville flow of Fig. 1.
This identification implies that, as a result of the above-mentioned
singular behaviour in the ultraviolet limit $A \rightarrow 0$,
the correlator ${\tilde A}_N$
cannot be interpreted as an $S$-matrix element, 
whenever there is a departure from criticality $s \ne 0$.

When one integrates over the Saalschultz contour in Fig. 1, the integration
around the simple pole at $A=0$ yields an imaginary part~\cite{kogan2,emn},
associated with the instability of the Liouville vacuum. We note, on the  
other hand, that the integral around the dashed contour
shown in Fig. 1, which does not encircle the pole at $A=0$, is well defined.
This can be intepreted as a well-defined $\nd{S}$-matrix element,
which is not, however, factorisable into a product of 
$S-$ and $S^\dagger -$matrix elements, due to the 
$t$ dependence acquired after the identification 
$t=-{\rm ln}A$. This formalism is similar to the 
Closed-Time-Path (CTP) formalism used in non-equilibrium 
quantum field theories~\cite{ctp}, as we now discuss. 

In the path-integral formulation of the CTP approach
to non-equilibrium field theory~\cite{ctp},
the partition function is expressed as an integral over the path of 
Fig. 2: 
\be
Z[J_1,J_2:\rho]=\int[D\Phi]<\Phi_1,t=t_0|\rho|\Phi_2,t=t_0>expi\{Tr\Phi D^{-1}\Phi + S_{int}[\Phi] + Tr J. \Phi \}
\label{ctppart}
\ee
where $[D\Phi]=[D\Phi_1][D\Phi_2]$, 
with 
$\Phi_1 (\Phi_2)$ denoting fields
whose time arguments are on the upper (lower) 
segments of the CTP, with corresponding actions 
$S_{int}[\Phi_{1,2}]$,
$D^{-1}$ is the inverse field propagator 
and $S_{int}[\Phi] = S_{int}[\Phi_1] - S_{int}[\Phi_2]$.
The density matrix in (\ref{ctppart}) may be 
represented as~\cite{ctp}:
\be
  <\Phi_1, t_0|\rho|\Phi_2,t_0>=exp\{\frac{i}{2}\int K[\Phi _j] \} 
\label{kernel2}
\ee
where the integral extends over space only, at some initial time 
$t_0$, and the kernel
$K[\Phi_i]$ may in general be expanded in powers of the 
boundary fields $\Phi_i$:
$ \int K[\Phi_j]= const + \int K_j\Phi_j + \int \frac{1}{2}\Phi_i K^{ij} \Phi_j
+ \int \frac{1}{6}K^{ijk}\Phi_i \Phi_j \Phi_k + \dots $.

\begin{figure}
\hglue4.0cm
\epsfig{figure=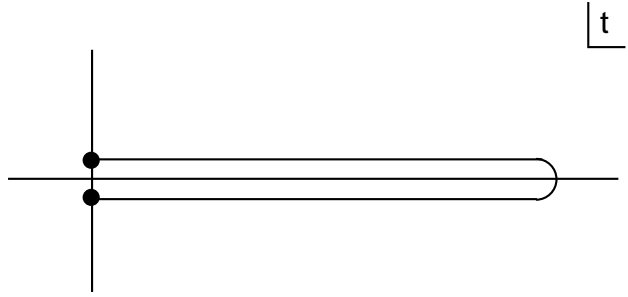}
\caption{ Contour of integration in the complex $t$ plane,
used in the Closed Time Path 
(CTP) formalism. In the Liouville string case $t=-{\rm ln}A$, where
$A$ is the world-sheet area.}
\end{figure}

In our string equivalent of this formalism, the r\^ole 
of the action $S_{int}[\Phi_i]$ is taken by the Zamolodchikov 
$C$ function (\ref{ctheorem}), which depends on the $\sigma$-model 
background couplings/fields $\{ g^i \}$. 
The topological summation over genera 
induces quantum fluctuations in the $\{ g^i \}$~\cite{emnd}
via a second-quantized effective partition function
$Z_{over genera} =\int [Dg]{\cal P}[g] e^{-C[g]}$:
$C[g]=c^* - g_s^\chi \int_{-\infty}^{+\infty} dt 
{\hat \beta}^i G_{ij} {\hat \beta}^j$, 
where $G_{ij}$ is the Zamolodchikov metric 
in theory space, which is occupied with a probability 
distribution ${\cal P}[g]$. An extreme case 
of a non-trivial topology leading to a divergence 
is one in which an infinitely long tube joins two distant 
parts of a Riemann surface. In this case, 
the string propagator along the tube takes the form
\be
{\cal D} =\sum_{m} \int_\delta \frac{dq}{q}\frac{d{\overline q}}{{\overline q}} 
q^{L_0} {\overline q}^{{\overline L}_0} |m><m|
\label{stringprop}
\ee
which has a double logarithmic divergence as the `size'
of the tube $\delta \rightarrow 0$. This scale may be 
absorbed in a generalized version 
of the  Fischler-Susskind mechanism~\cite{fs},
by introducing quantized couplings, 
${\hat g}^i(t): t=-{\rm ln}\delta$, with a theory-space
probability distribution 
\be
   {\cal P}[g] = 
e^{-({\hat g}^i(t)- g^{*i})G_{ij}({\hat g}^j(t)-g^{*j})/\Gamma (\delta) }
\label{prob}
\ee
where 
$\Gamma(\delta) \propto |{\rm ln}\delta|$ as $\delta \rightarrow 0$ 
absorbs the 
infinities associated with 
the pinched world-sheet tube. 
The analysis requires the introduction
of world-sheet wormhole parameters~\cite{recoil,emnd,lizzi},
which
parallels the treatment of 
space-time wormhole parameters
in four-dimensional field theories~\cite{coleman}.
In our interpretation of the Liouville scale as target time,
flowing opposite to the conventional 
renormalization-group flow~\cite{emn,kogan,kogan2}, 
the above picture matches the standard renormalization-group
approach, when one identifies 
$t_\infty$ with  
the infrared fixed point. In this way one arrives at a 
(Liouville) string 
equivalent of the initial-state density matrix  (\ref{kernel2}). 

After these preliminaries, we are now ready to present our construction 
of $D$-brane foam, as an example of the above formalism. 
Compared to previous quantum treatments of $D$ branes~\cite{dbranes}, the key physics step is to find a representation of the appearance 
and disappearance of virtual $D$ branes. 
We shall argue that this may be found in the context 
of the $\sigma$-model description of point defects
on the world sheet, namely monopoles 
and vortices, which are described 
by the following partition function~\cite{ovrut}:
\bea
Z&=&\int D{\tilde X} exp(-\beta S_{eff}({\tilde X}) )  \nn \\
\nonumber
\beta S_{eff}&=&  \int d^2 z [ 2\partial {\tilde X}
{\overline \partial } {\tilde X} +  
\frac{1}{4\pi }
[ \gamma _v\omega ^{\frac{\alpha}{2}-2}
(2 \sqrt{|g(z)|})^{1-\frac{\alpha}{4}}: cos (\sqrt{2\pi\sqrt{\beta}q }
[{\tilde X}(z) + {\tilde X}({\bar z})]):   \\
& +&   
\frac{1}{4\pi }
[ \gamma _v\omega ^{\frac{\alpha'}{2}-2}
(2 \sqrt{|g(z)|})^{1-\frac{\alpha'}{4}}: cos (\frac{e}{\sqrt{\beta}}
[{\tilde X}(z) - {\tilde X}({\bar z})]):]  
\label{sevenv}
\eea
where 
$\alpha \equiv 2\pi\beta q$,$\alpha' \equiv e^2/2\pi\beta$:
$q$ ($e$) is the vortex (monopole) charge, $\gamma_{v,m}$ are the 
fugacities for vortices 
and monopoles,  
and 
${\tilde X}$ is a $\sigma$-model field, 
whose 
world-sheet equation of motion admits 
vortex and monopole solutions: 
\be
\partial _z\partial _{\bar z} {\tilde X}_v =i\pi \frac{q}{2}
[\delta (z-z_1)-\delta(z-z_2)],  \qquad 
   \partial _z  \partial _{\bar z} {\tilde X}_m =-\frac{e \pi} {2}
[\delta (z-z_1) -\delta (z-z_2)]
\label{sourcev}
\ee
We note that $\beta^{-1}$ plays the r\^ole of an effective 
temperature in (\ref{sevenv}), which requires for its specification
an ultraviolet (angular) cut-off $\omega$. The vortex and monopole 
operators have anomalous dimensions: 
\be
      \Delta _m =\frac{\alpha }{4}=
      \frac{\pi\beta}{2}q^2, 
\qquad 
   \Delta _m =\frac{\alpha '}{4}=
      \frac{e^2}{8\pi\beta}
\label{confdim}
\ee 
and the system (\ref{sevenv}) 
is invariant under the T-duality transformation~\cite{ovrut}: 
$\pi \beta \leftarrow\rightarrow \frac{1}{4\pi\beta}~;~q 
\leftarrow\rightarrow e$. 
We identify ${\tilde X}$ in (\ref{sevenv}) 
with the rescaled Liouville field $\sqrt{\frac{C-25}{3g_s^\chi}}\phi$,
in which classical solutions to the equations of motion and  
spin-wave fluctuations have been subtracted~\cite{ovrut}, 
and relate $\beta$ to the 
Zamolodchikov $C$ function of the accompanying matter~\cite{ovrut}: 
$\beta \rightarrow \frac{3g_s^\chi}{\pi(C-25)}$. 
We will be interested in the case $C > 25$, and 
we shall consider only irrelevant deformations,
that do not 
drive the theory to a 
new fixed point. The world-sheet  
system is then in a dipole phase~\cite{ovrut}.
We note that this system has the general 
features discussed previously, namely non-conformal 
deformations,
non-trivial Liouville dynamics, and dependence 
on an ultraviolet cutoff. Furthermore, this system 
was discussed previously~\cite{emn} as a model 
for space time foam in the context of $(1+1)$-dimensional
string black holes. 

To see why we claim that it can be used 
to represent
$D$-brane foam, we first  
consider the scattering of close string states
$V_T(X)=exp(ik_M X^M - iEX^0): 
M=1, \dots D_{cr}-1$, where $D_{cr}$ is the 
critical space-time dimension, in the presence 
of a monopole defect. An essential aspect of this
problem is the singular behaviour of the operator product
expansion of $V_T$ and a monopole operator $V_m$. Treating
the latter as a sine-Gordon deformation of (\ref{sevenv}),
computing at the tree level using the free world-sheet
action, and suppressing for brevity anti-holomorphic parts,
we find
\bea
&~&\hbox{Lim}_{z \rightarrow w}<V_T (X^0,X^i)(z) 
V_{m} (X^0)(w) \dots > \nn \\
&~&\sim \int d^Dk \int dE \delta^D (k) 
(\dots) 
[\delta (\Sigma E + \beta ) (z-w)^{-\Delta _T - \Delta _m} + 
\nn \\
&~&\delta (\Sigma E - \beta ) (z-w)^{-\Delta _T - \Delta _m} ] 
\label{twopoint}
\eea
where $\Delta _T =\frac{E^2}{2}$, 
the energy-conservation $\delta$ functions result from integration
over the Liouville field ${\tilde X} \equiv X^0$, and
$(\dots)$ indicates factors 
related to the spatial momentum components of 
$V_T$, other vertex operators in the correlation function, etc..  
We see that (\ref{twopoint}) 
has cuts for generic values of 
$\Delta_T + \Delta _m$, causing the theory 
to become that of an open string.

In the particular case of the tree-level three-point correlation 
function 
$   {\cal A}_{TTm} = <V_T(E_1)V_T(E_2)V_m(e/\beta^{1/2})> $,
integration over the Liouville field $X^0$ imposes 
energy conservation in the form:
${\cal A}_{TTm} \propto 
[\delta (E_1 + E_2 + e/\sqrt{\beta})
{\cal A}_1' 
+ \delta (E_1 + E_2 - e/\sqrt{\beta}){\cal A}_2']$
where the two terms arise from the cosine form of the 
monopole vertex operator (\ref{sevenv}), though only 
the second term is in the physical region $E_{1,2} >0$: 
\be
E_1 + E_2 =  e/\sqrt{\beta} = e\left(\pi (C -25)/3g_s^\chi\right)^{\frac{1}{2}}
\label{energycons}
\ee
This equation is consistent with the monopole describing a massive
particle 
in target space-time if $C > 25$. For this, we consider 
a model with 25 additional space-like coordinates $X^i$ coupled to a linear 
dilaton background~\cite{aben}: 
$\int d^2z \partial X^i \partial X^j \delta _{ij}+ 
i\int d^2z Q\eta_i X^i R^{(2)}$, where $\eta_i$ a fixed vector, 
which has $C=26 + {\cal O}(\beta^{i})^2$
if $Q=\sqrt{{C - 25 \over 3}}$.

In the open string picture (\ref{twopoint}),  
induced by the interaction of matter with the defects,
one may assume that the world-sheet curvature $R^{(2)}$ 
is concentrated
along the cut, in which case  
the linear-dilaton background term  
contributes an extrinsic-curvature term ${\hat k}$ on the world-sheet boundary 
\be 
i\int_{\partial \Sigma} Q\eta_i X^i {\hat k}   
\label{lineardilaextr}
\ee
Since we wish to study the effects of virtual $D$ branes,
we need to be able to sew tree-level amplitudes into loop
diagrams with internal monopoles. This requires a better
understanding of the appropriate boundary conditions
along the cut in the world sheet. 
In the case of the Liouville field $\phi=X^0$, the 
appropriate boundary condition in open string theory has been 
derived in \cite{ambjorn}:
\be
  \partial _n X^0  + i\frac{{\hat k}}{2} Q =0 
\label{genericboundary}
\ee
where ${\hat k}$ is the extrinsic curvature of the fiducial metric, 
up to a possible cosmological constant term and higher-order 
quantum corrections. We note that (\ref{genericboundary}) 
reduces to Neumann boundary conditions 
in the critical limit $Q \rightarrow 0$, as is the case in the simplified 
situation considered here and
in ref. \cite{ovrut}, where 
the cosmological constant term is ignored in  
the Liouville action, thereby allowing the decoupling 
of the spin-wave part from the monopole field ${\tilde X} \equiv X^0$.  
The space coordinates
$X^i$ may be taken to have either Neumann or Dirichlet 
boundary conditions, which are related by a T-duality transformation. 
If Neumann boundary conditions are chosen, one must introduce 
a background gauge field $A_i$ related to possible Chan-Paton factors 
at the ends of the open string, and the corresponding 
$\sigma$-model path integral is: 
\be
Z=\int DX^i DX^0 e^{-\int _{\Sigma} d^2z \frac{1}{4\pi\alpha '} [\partial X^i {\overline \partial}
X^i\delta _{ij} - \partial X^0 {\overline \partial}X^0] 
- i\int _{\partial \Sigma} A_i(X^0) \partial _\tau X^i + i\int _{\partial \Sigma} {\hat k} Q \eta_iX^i} 
\label{opensigma}
\ee
where 
$\partial_\tau$ denotes a tangential derivative on the 
world-sheet boundary $\partial \Sigma$. 
The Abelian background gauge field 
depends on time $X^0$ only, and we work in the gauge 
$A_0 = 0$ for simplicity. 

It is convenient
for our purposes to make a $T$-duality transformation:
$Y_\alpha^i=\partial_\alpha X^i$: 
$\epsilon_{\alpha\beta}\partial_\alpha Y_\beta^i=0$,
which yields a tractable weak-coupling formalism. This we
implement in (\ref{opensigma}) using Lagrange multipliers
$\lambda_\alpha^i, {\widehat X}^i$~\cite{otto}:
\bea
&~& Z=\int D{\widehat X}^i \int DX^i 
DY^i_\alpha \delta (Y^i_\alpha -\partial_\alpha X^i)
e^{-\int_\Sigma d^2z [(Y_\alpha ^i)^2 
-\partial X^0 {\overline \partial}X^0]  - i\int _{\partial \Sigma} 
A_i(X^0) Y_\tau ^i} \nn \\  
&~&e^{- i\int _\Sigma d^2z {\widehat X}^i 
\epsilon_{\alpha\beta}\partial_\beta Y^\alpha_i 
-i\int _{\partial \Sigma} 
{\hat k} Q\eta_i X^i} 
= \nn \\
&~& \int D{\widehat X}^i \int D\lambda_\tau ^i \int DX^i DY^\alpha_i 
e^{-\int_\Sigma d^2z [(Y_\alpha ^i)^2 
-\partial X^0 {\overline \partial}X^0] -  i\int _{\partial \Sigma} 
A_i(X^0) Y_\tau ^i -i\int_{\partial  \Sigma} Y_\tau^i {\widehat X}^i} \nn \\
&~&e^{i\int _\Sigma d^2z Y_\alpha^i 
\epsilon_{\alpha\beta}\partial_\beta {\widehat X}^i
- i\int_{\partial \Sigma} 
(Y_\tau^i \lambda_\tau^i + X^i\partial _\tau \lambda_\tau^i + 
{\hat k}Q\eta_iX^i )}
\label{pixtilde}
\eea
The boundary term in the integration over $Y_\tau^i$ 
results in the Dirichlet constraints: 
\be
{\widehat X}^i + \lambda _\tau ^i + A^i(X^0) =0,
~~\partial_\tau \lambda_\tau^i + i{\hat k} Q \eta^i =0 \qquad {\rm on~\partial \Sigma} 
\label{dirichlet}
\ee
with ${\widehat X}^i$ described by a free $\sigma$-model action
in the bulk~\footnote{This has 
Euclidean signature: the appropriate Minkowski theory is obtained by
analytic continuation.}. The boundary condition (\ref{dirichlet})
is not conformally invariant, which is known to be the case
for fixed Dirichlet boundary conditions 
in the presence of a linear-dilaton background~\cite{dbranes}.
This `conformal anomaly' in the dual theory
keeps track of the fact that the matter system has central charge 26,
but a non-critical number ($D=25$) of spatial  
dimensions~\cite{aben}. 

Conformal invariance can be restored
by Liouville dressing~\footnote{Some authors have advocated restoration 
of conformal invariance for the Dirichlet 
linear-dilaton system by modifying 
the standard $D$-brane boundary state
by imposing appropriate boundary interactions~\cite{LiD}, which 
would restore conformal invariance for the matter system alone.
However, here we implement Liouville dressing of the 
non-critical matter system, which is crucial 
for our interpretation of time as a Liouville mode~\cite{emn,aben,kogan}.}. 
Fixed Dirichlet boundary conditions are obtained by shifting 
the $X^0$ field in such a way so that 
$A_i(X^0)+ \lambda_\tau^i=A_i'(X^{'0})$, which
can be done with an appropriate choice of $A(X^0)$ (`Liouville 
dressing'). The simplest choice is a gauge potential 
corresponding to a constant `electric' field $v_i$: 
\be
   A_i(X^0)={\rm const} +  v_iX^0
\label{electricgauge}
\ee
where 
$v_i$ is determined from (\ref{dirichlet}), and is found to be 
proportional to $\eta_i {\hat k}Q$. 
Physically, this means that a non-critical-dimension 
matter string theory, in a 
linear-dilaton background~\cite{aben}, 
corresponds, upon $T$-dualization in our Liouville framework, 
to a moving $D$-brane with a velocity $v_i$ determined by  
the matter central charge deficit $Q$.
The `motion' of the $D$-brane implies target-time dependence 
provided by the Liouville field~\cite{emn}.  

We note that, by applying the above dualization procedure
for $p$ of the spatial coordinates,  one can 
generate generic $p$-brane configurations, with $p \ge 1$ Dirichlet 
directions.
The $D0$ particle is the first non-trivial structure in this 
hierarchy of stringy structures.
Since the collective coordinates of such $D0$ branes 
are associated with a canonically-quantized phase space~\cite{lizzi,emnd}, 
the above construction may be thought of as providing for the 
emergence of a target space-time from string solitonic structures, 
where 
the space is provided by the collective coordinates
describing the position of the string soliton, and the time 
is given by the 
Liouville mode of the non-critical 
two-dimensional model describing the interaction of string matter 
with
the world-sheet defects.

{}From the point of view of the target-space wave function of the string, 
the above background gauge fields appear as Bohm-Aharonov phase factors. 
To see this, suppose that a closed-string state encounters a monopole defect 
on the world sheet at $(\sigma_0,\tau_0)$, corresponding to 
a spatial location $y^i$ and a time
$X^0=\tau_0$ in the light-cone gauge. The gauge field was absent before
this encounter, 
so has a step-function singularity:
\be
A(X^0)^i=(y^i-(X^0-\tau_0)F^{0i})\Theta(X^0-\tau_0) 
\label{thetaelectric}
\ee 
where $F^{0i}$ is the electric field strength, which has a
$\delta (X^0-\tau_0)$ 
singularity, but is otherwise conformally invariant. In the T-dual picture, there is a corresponding 
`sudden' appearance of Dirichlet  
boundary conditions, breaking conformal invariance, that we interpret as the excitation of a 
$D$ brane in the vacuum. There is a corresponding singularity in the
space-time 
curvature~\cite{diffusion} associated with the creation of the world-sheet 
monopole: 
\be 
     R~\ni~-2\frac{25\delta (X^0-\tau_0) }{[1 + \Theta(X^0-\tau_0)
(\sum_{i=1}^{25} y_i^2)]^2} 
\label{curvature}
\ee
which 
confirms our interpretation that the appearance 
of the world-sheet monopole corresponds to the
appearance 
of a black hole represented by a $D$ brane. 

Propagation of a light closed-string particle through this representation
of $D$-brane foam
involves, at the lowest order, a diagram with a disk topology, internal tachyon vertices, and 
the boundary conditions (\ref{genericboundary}, \ref{dirichlet}). This may describe 
scattering through a real (or virtual) $D$-brane state, with production
and decay amplitudes
$A_{TTm}$, subject to the energy-conservation condition (\ref{energycons}). 
The next term in a topological expansion in genus
$g=2-2\#_{handles}-\#_{holes}$ is
an annulus with closed-string operator insertions. As has been discussed elsewhere~\cite{periwal},
this has a singularity 
${\cal A} \sim  \delta (E_1+E_2)\sqrt{\frac{1}{{\rm ln}(\delta )}}$
in the pinched annulus configuration $\delta \rightarrow 0$, which is regularized 
by introducing recoil 
operators~\cite{emnd,kogwheat,diffusion} $C,D$ 
to describe the back reaction 
of the struck $D$ brane:  
\be
 V_{rec} = y_i C + u_i D \qquad : 
C \equiv \epsilon \int _{\partial \Sigma} \Theta_\epsilon (X^0) 
\partial_n X^i 
\qquad D  \equiv \int _{\partial \Sigma} X^0 \Theta_\epsilon (X^0) 
\partial_n X^i 
\label{CDpair}
\ee
where ${\rm lim}_{\epsilon \rightarrow 0}\Theta _\epsilon (X^0)$ is a suitable 
integral representation of the step function, and $y_i$, $u_i$ are the position 
and momentum of the recoiling $D$ brane. As discussed in~\cite{diffusion}, we
identify $1/\epsilon^2 \sim \hbox{ln} \delta$, and in turn, using the
Fischler-Susskind mechanism~\cite{fs} 
on the world sheet to relate renormalization-group infinities 
among different genus surfaces, we identify $t \sim {\rm ln}\delta$.   
The operators $C, D$ 
consitute a logarithmic pair~\cite{kogwheat} with 
$<C(z)C(0)>$, $<C(z)D(0)>$ 
non-singular as $\epsilon \rightarrow 0^+$, whereas   
$<D(z)D(0)>$ is singular with a world-sheet 
scale dependence~\cite{kogwheat} $D \rightarrow D + 
tC$, from which we infer that $u_i \rightarrow u_i$,
$y_i \rightarrow y_i + u_i t$, corresponding to a Galilean 
time transformation~\cite{lizzi}, as is appropriate 
for a heavy $D$ brane with mass $\propto 1/g_s$ (\ref{velocityoptimum}). 

The logarithmic operators (\ref{CDpair}) 
make divergent contributions  to the genus-0 amplitude 
in the limit where it becomes a pair of Riemann surfaces 
$\Sigma_1, \Sigma_2$ connected by a degenerate strip~\cite{emnd,lizzi}:  
\bea
{\cal A}_{strip} \sim &g_s& ({\rm ln}^2 \delta) \int d^2 z_1 D(z_1)
\int d^2 z_2 C(z_2), \nn \\ 
 &g_s& {\rm ln}\delta\int d^2 z_1 D(z_1) \int d^2 z_2 D(z_2), \quad 
g_s{\rm ln}\delta \int d^2 z_1 C(z_1) 
\int d^2 z_2 C(z_2)
\label{stripdiv}
\eea
Assuming a dilute gas of monopole defects on the world sheet, the amplitudes
(\ref{stripdiv}) become contributions to the effective action~\cite{emnd,lizzi}. 
One may then seek to cancel them or else to absorb them into scale-dependent 
$\sigma$-model couplings as described in equations 
(\ref{stringprop}, \ref{prob}). If they could all be cancelled, 
the corresponding $\sigma$-model would be conformally invariant, 
whereas absorption of these divergent contributions would result 
in departures from criticality. 

The leading double logarithm 
associated with the CD combination in (\ref{stripdiv}) 
may indeed be cancelled~\cite{lizzi} by imposing the momentum conservation 
condition
\begin{equation}
u_i =  g_s(k_1 + k_2)_i  
\label{velocityoptimum}
\end{equation}
as expected for a $D$ brane soliton of mass $1/g_s$, which 
is also consistent with the tree-level energy-conservation 
condition in (\ref{energycons}), obtained by integration 
over the Liouville zero mode. 
{}From the point of 
view of the Liouville theory on the open world sheet 
(\ref{pixtilde}, \ref{dirichlet}), the tree-level monopole 
mass term arises from a boundary term $i\int_{\partial \Sigma}
QX^0{\hat k}$ in the effective action~\cite{ambjorn}, where
${\hat k}$ is the extrinsic curvature 
and $Q$ is given by equation (\ref{31/2}). 
When the Liouville 
integration is performed at the quantum level, $Q$ is replaced
by its value in (\ref{31/2}), which receives
a contribution from $\beta_{y_i}=u_i$~\cite{diffusion}, by virtue of the 
logarithmic operator product expansion of $C$ and $D$~\cite{kogwheat}. 
Expanding the right-hand-side of (\ref{31/2}), using (\ref{ctheorem}),   
for small $|u_iu^i| << 1$,  
we find the quantum energy conservation condition:
\bea
E_1 + E_2 =\frac{e}{\sqrt{\beta}}=e\left(\pi(C-25)/3g_s\right)^{1/2}
= \frac{e}{\sqrt{g_s}}(1 + \frac{u_i^2}{2} + \dots )
\label{qecons}
\eea
The result (\ref{qecons})
matches the momentum conservation condition (\ref{velocityoptimum})
upon setting $e\sqrt{\pi/3}=1/\sqrt{g_s}$, thereby  
confirming our interpretation of time as the Liouville field~\cite{emn,kogan}. 
Thus, cancellation of the leading double logarithm 
in (\ref{stripdiv}) enforces energy conservation $\frac{d}{dt}<E>=0$,
as argued previously~\cite{emn} in a general renormalization-group approach
to Liouville dynamics. 

The single logarithms associated with the CC and DD contributions 
in (\ref{stripdiv}) are a different story, since they 
can only be absorbed into quantum coupling parameters~\cite{emnd,lizzi}:
${\hat y_i}=y_i + \alpha_C\sqrt{{\rm ln}\delta}, {\hat u}_i = u_i + 
\alpha_D\sqrt{{\rm ln}\delta}$. 
The resulting probability distribution in theory space (\ref{prob}) 
becomes time dependent~\cite{emnd,lizzi}:
\begin{equation}
{\cal P} \sim 
 \frac{1}{g_s^2ln\delta} 
e^{-\frac{({\hat q}_m -q_m)G^{mn}({\hat q}_n - q_n)}{g_s^2ln\delta}} 
\label{timprob}
\end{equation}
where $q_m \equiv \{ y_i,    u_i \}$. 
Within the CTP-like interpretation of the Saalschutz contour reviewed earlier,
this corresponds to a time-dependent $\nd{S}$-matrix transition from the 
initial-state density matrix. 

Although this is compatible with energy conservation in the mean, 
as discussed previously, it entails a non-quantum-mechanical 
modification of the energy fluctuations $(E-<E>)^2$. It has been shown 
elsewhere~\cite{emn} that these may be related to the non-zero renormalization
group functions $\beta_i$:
\be
\frac{d}{dt}<(E-<E>)^2> = <(\frac{d}{d t}\beta^i) G_{ij} \beta^j> 
\label{energfluct}
\ee
It has been shown~\cite{emnd} in the Liouville-string framework
that $\frac{d}{dt}G_{ij}=QG_{ij}$. Thus, using the result~\cite{emn}
that ${\dot S} =\beta^iG_{ij}\beta^j$, we find
\be
  \frac{d}{dt}<(E-<E>)^2> = \frac{1}{2}{\ddot S}-Q{\dot S}
\label{energ-entr}
\ee 
which may be considered as 
a quantum-gravitational version of the 
fluctuation-dissipation theorem
of statistical mechanics~\cite{emn,emnd}. We see that,
the closer the system lies to its 
infrared fixed point as $t \rightarrow \infty$, 
the more classical 
it becomes, in the sense of an increase in entropy 
and a corresponding decrease in its energy fluctuations.
This result in the context of Liouville $D$ branes
is in  agreement with the general picture, advocated in \cite{emn},
that a classical field-theoretic vacuum is obtained from a non-critical 
string theory via decoherence in `theory space'. 

To acquire some feeling for the possible order of magnitude of such
decohering effects, we estimate from (\ref{velocityoptimum}) that
$|\beta_{y_i}|=|u_i|={\cal O}(E/M_P)$, where $E$ is the typical energy 
of a closed-string light-particle state, and we neglect numerical factors, 
powers of $g_s$, etc. Correspondingly, assuming an effective 
$D$-brane density of order unity per Planck volume, we estimate ${\dot S}
= {\cal O}(E^2/M_P^2)$, whereas 
$\frac{d}{dt}<(E-<E>)^2>$ vanishes to this order,
since our heavy branes do not 
accelerate: $\frac{d}{dt} u_i =0$~\cite{diffusion,lizzi}. 
As has been discussed elsewhere~\cite{emn}, the corresponding 
time-evolution equation for the density matrix 
$\rho$ of a light-particle state takes the form:
\be
\frac{\partial}{\partial t} \rho = i [H, \rho ] + \nd{\delta H}\rho 
: ~~\nd{\delta H}=i\beta^i G_{ij}[g^j,~]
\label{liouville}
\ee
and we estimate $|\nd{\delta H}|={\cal O}(E^2/M_P)$, in agreement 
with previous estimates~\cite{emn,emnw}, and close in order of magnitude
to the experimental bound from the $K^0-{\overline K}^0$
system~\cite{ehns,CPLEAR}. 
In higher orders, we expect 
$\frac{d}{dt}\beta^i = \beta^m \partial_m \beta^i \sim 
2\beta^m~(c^i_{mk}g^k + {\cal O}[g^2]) $, 
where the operator-product-expansion
coefficients $c^i_{jk}$ are of order $E^2/M_P^2$ in the closed-string sector,
or $E/M_P$ in the open-string sector, as in the case of $D$ branes. 
Thus, (\ref{energfluct}) is suppressed -  as compared to ${\nd{\delta H}}$
(\ref{liouville}) -   
by higher powers of Planck Mass,
at least as $E^3/M_P^2$ in the open-string sector, which makes 
such energy fluctuations difficult to detect in foreseeable
experimental facilitites. 

Finally, we comment on the
new uncertainty relations that stem from the above 
construction, which could be used to 
probe the quantum-gravity structure of a $D$-brane space time. 
The key observation is that 
the target time  $X^0 
\propto \sqrt{C-25}\varphi$, where $\varphi$ is the 
original Liouville field
appearing in the conformal scale factor of the world-sheet 
metric. 
{}From (\ref{31/2}), then, and the fact that summing over 
world-sheet genera leads to a canonical quantization of 
the $\sigma$-model couplings 
$y_i,u_i$~\cite{emnd,lizzi},
we see that in this picture $X^0$ appears as an 
`operator' in the $D$-brane collective phase space, leading to 
non-trivial commutation relations between the time $X^0$
and the position (collective) coordinates 
of the $D$ brane~\cite{aemn}. It can be easily seen that, for 
slowly-moving non-relativistic branes: $|u_i|^2 << 1$ as we are
considering 
here, such commutators  
lead to the following space-time 
uncertainty relation:
\be 
    \Delta y_i \Delta X^0 \ge g_s^{1/2} u_i + \dots 
\label{uncert}
\ee
Non-trivial space time uncertainty relations of the form (\ref{uncert}),
but {\it independent} of the string coupling, 
had been derived previously in the context of {\it critical} 
(conformal) strings and $D$ branes
in ref. \cite{yoneya}.

\end{document}